\documentclass[aps,prl,preprint,groupedaddress]{revtex4-1}
\usepackage{siunitx}%si単位
\usepackage{graphicx}
\usepackage{amsmath}
\begin{document}

%\preprint{}

\title{In-situ tuning of whispering gallery modes of levitated silica microspheres}


\author{Yosuke Minowa}
\email[]{minowa@mp.es.osaka-u.ac.jp}
\author{Yusuke Toyota}
\author{Masaaki Ashida}
%\thanks{}
\affiliation{Graduate School of Engineering Science, Osaka University, Toyonaka, Osaka 560-8531, Japan}


\date{\today}

\begin{abstract}
We demonstrated the tuning of whispering gallery modes (WGMs) of a silica microsphere during optical levitation through the annealing process. We determined the annealing temperature from the power balance between the CO2 laser light heating and several cooling processes. Cooling caused by heat conduction through the surrounding air molecules is the dominant process. We achieved a blue shift of the WGMs as large as 1 \%, which was observed in the white-light scattering spectrum from the levitated microsphere.
\end{abstract}


%\keywords{}

\maketitle


\section{Introduction}

Optomechanical response of a single levitated solid microsphere to light is largely affected by whispering gallery modes (WGMs) excitation\cite{ashkin_observation_1977}. Therefore, precise tuning of the WGMs is necessary, as the radiation pressure can be resonantly enhanced through WGM excitation\cite{barker_doppler_2010,li_cooling_2014,li_simultaneous_2016}. The optomechanical effect on the levitated micro and nanospheres is at the core of the active or passive cooling of the center of motion of the levitated particles\cite{romero-isart_toward_2010,chang_cavity_2010,li_millikelvin_2011,jain_direct_2016,vovrosh_controlling_2016} down to the quantum ground state, which would lead to the macroscopic quantum phenomena\cite{romero-isart_quantum_2011,rashid_experimental_2016}. Furthermore, the exploration of optical binding effect and optomechanical coupling between multiple microshperes\cite{ng_strong_2005,povinelli_high-q_2005,yang_spectroscopy_2009,arita_rotation_2015,kudo_optical_2016} requires mutual tuning of the WGMs. Since it is impossible to levitate an exact predetermined particle with commonly used loading techniques\cite{li_millikelvin_2011,neukirch_observation_2013,monteiro_dynamics_2013,minowa_optical_2015}, in-situ modification of the particle size or refractive index is needed to tune the WGMs.

Solid microspheres are one of the most intriguing microcavity structures, which could have WGM resonances with an extremely high quality factor and small mode volume due to the three dimensional light confinement \cite{spillane_ultralow-threshold_2002}. The high quality factor and small mode volume of the solid microspheres lead to cavity quantum electrodynamic phenomena\cite{buck_optimal_2003}, ultra-low threshold lasing\cite{okamoto_fabrication_2014}, and highly efficient optical nonlinearity\cite{treussart_evidence_1998}. Liquid microspheres or microdroplets also show excellent cavity performance due to naturally smooth surface and high-sphericity\cite{kiraz_volume_2007,schafer_quantum_2008}. The size and shape of the microdroplet can be controlled through evaporation\cite{kiraz_volume_2007} or coagulation\cite{hopkins_control_2004} and radiation pressure\cite{aas_spectral_2013}, leading to the tuning of the WGM resonances. The WGM tuning of the solid microspheres, however, continues to be a challenge owing to the technical difficulties that stems from the stable and strong atomic bonding. One of the most reliable methods is the temperature control through the thermal expansion and the temperature dependent refractive index\cite{chiba_resonant_2004}. Although the method provides the precise and repeatable tuning of WGMs, the requirement of the temperature stabilization prevents its use in many situations including optical levitation. 

Here, this study reports on the in-situ tuning of WGMs in levitated silica microspheres through the annealing process. The annealing of silica affects the microscopic silica network structure and the amount of impurity, leading to changes in the optical properties and the volume of the silica sample\cite{kaiser_drawing-induced_1974,cho_stabilization_2014}. Therefore, we can achieve different WGM resonances after different annealing conditions. The annealing was performed with the CO2 laser irradiation on the levitated silica microsphere. CO2 laser light with the wavelength of \SI{10.6}{\micro \meter} resonantly excites the broad Si-O-Si bands and the strong absorption occurs. The thermal balance between CO2 laser heating and several different cooling processes determines the annealing temperature. Thus, we can tune the WGMs by changing the CO2 laser irradiation power. We identified the WGM resonances from white light scattering spectra as the scattering cross section of the microspheres reflects the WGMs\cite{bohren_absorption_1983, schietinger_coupling_2009}.

\section{Experimental setup}
\label{sec:setup}

\begin{figure}[htbp]
\centering
\includegraphics[width=0.8\linewidth]{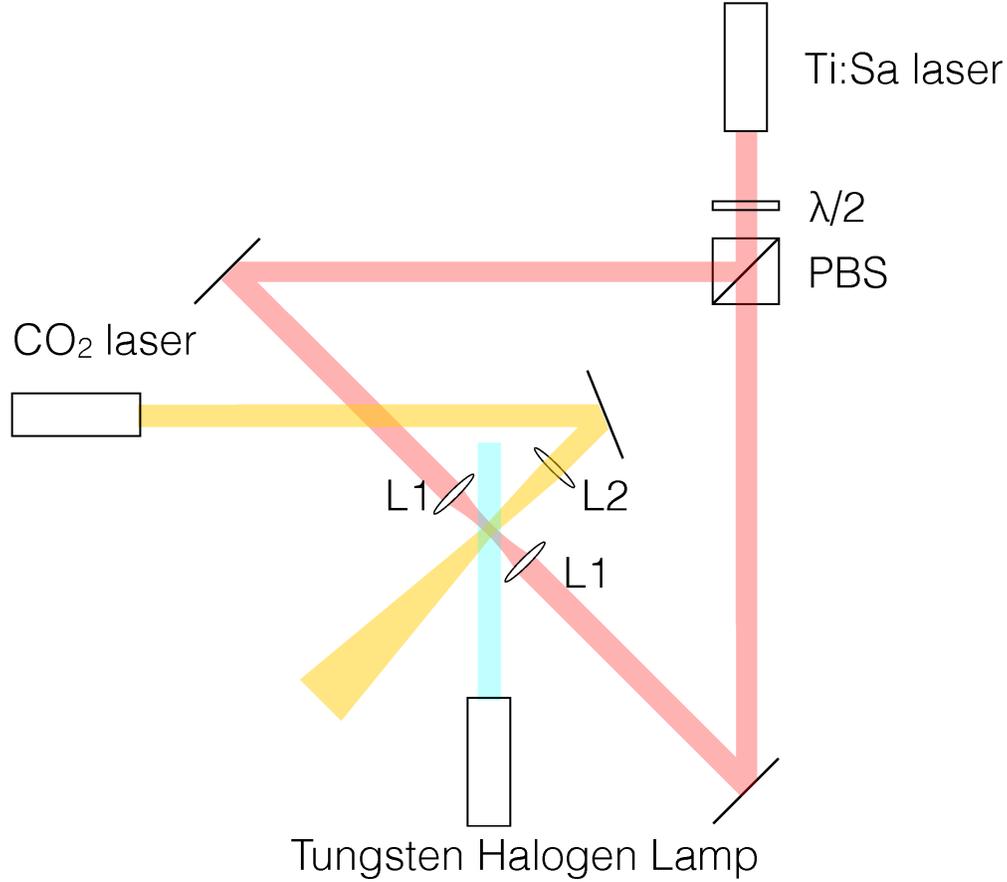}
\caption{Schematic representation of the dual beam optical trap setup for in-situ tuning of WGM resonances. Light from a continuous-wave Ti:Sapphire laser was split into two beams using a PBS. The power ratio between the two beams was controlled by rotating a half wave plate. Each beam was tightly focused by an aspherical lens (L1) with a common focal point forming a optical trapping potential. Optically levitated microspheres can be illuminated by CO2 laser light or white light from tungsten halogen lamp. We measured the scattering spectrum from a single optically levitated microsphere using a spectrograph.}
\label{fig:schematic}
\end{figure}

We optically levitated 3-\si{\micro \meter}-diameter microspheres (Polysciences, Silica Microspheres, 24330) with a dual-beam optical trap by using long-working-distance aspheric lenses to have enough space for CO2 laser irradiation (Fig. \ref{fig:schematic}). A continuous-wave Ti:Sapphire laser beam with the wavelength of \SI{785}{nm} was split into two parts using a polarizing beam splitter (PBS). The half-wave plate before the PBS was rotated to ensure equal power ($\sim$\SI{1}{W}) in the two beams. The two identical aspheric lenses (thorlabs, C240TME-B) focused the beams onto the same point, which leads to a cancelation of the scattering forces\cite{li_millikelvin_2011}. Silica microspheres were loaded into the trapping sight via an ultrasonic nebulizer (Omron, NE-U22)\cite{neukirch_observation_2013,monteiro_dynamics_2013,minowa_optical_2015}. \SI{100}{\micro\litre} of the stock solution was dried, redissolved in \SI{15}{\milli \litre} ethanol, and was sonicated in an ultrasonic bath for 20 minutes before use.

From a single optically levitated silica microsphere, we measured the white light scattering spectrum. The spectrum represents the WGMs in the microsphere according to Mie scattering theory\cite{bohren_absorption_1983,schietinger_coupling_2009}. Then, the optically levitated microsphere was heated with the CO2 laser irradiation for 2 minutes. The CO2 laser light was focused onto the levitated microsphere with a ZnSe lens of 100 mm focal length. After switching off the CO2 laser light, we measured the white light scattering spectrum again. The spectrum reflects the change in the WGMs due to the annealing. We repeated the process with different CO2 laser powers, i.e., different annealing conditions.   

\section{Results and discussion}
\label{sec:result}
\begin{figure}[htbp]
\centering
\includegraphics[width=0.8\linewidth]{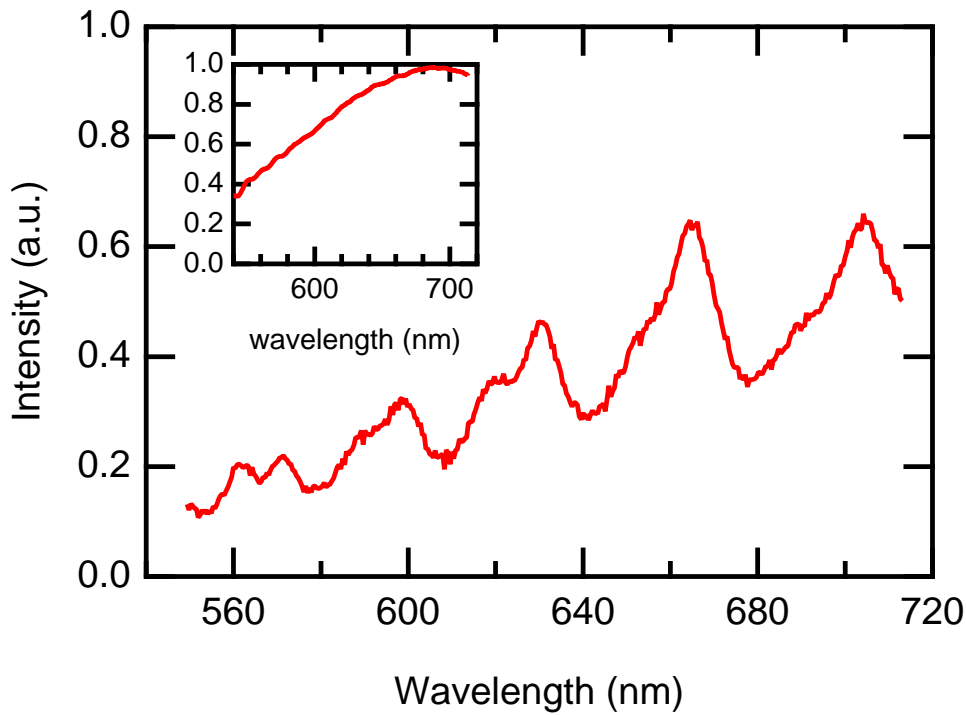}
\caption{White light scattering spectrum obtained from a single levitated microsphere. Each peak corresponds to a particular whispering gallery mode resonance. The inset shows the original white light source spectrum.}
\label{fig:wgm}
\end{figure}

Figure \ref{fig:wgm} illustrates the white light scattering spectrum from the levitated silica microsphere. The widths of the resonances were determined by several loss mechanisms including intrinsic or impurity material absorption loss, tunneling loss, and surface absorption or scattering loss. The original white light source spectrum shows a gradual increase toward longer wavelength as depicted in the inset of Fig. \ref{fig:wgm} and well describes the different peak heights for different resonances. 

\begin{figure}[htbp]
\centering
\includegraphics[width=0.8\linewidth]{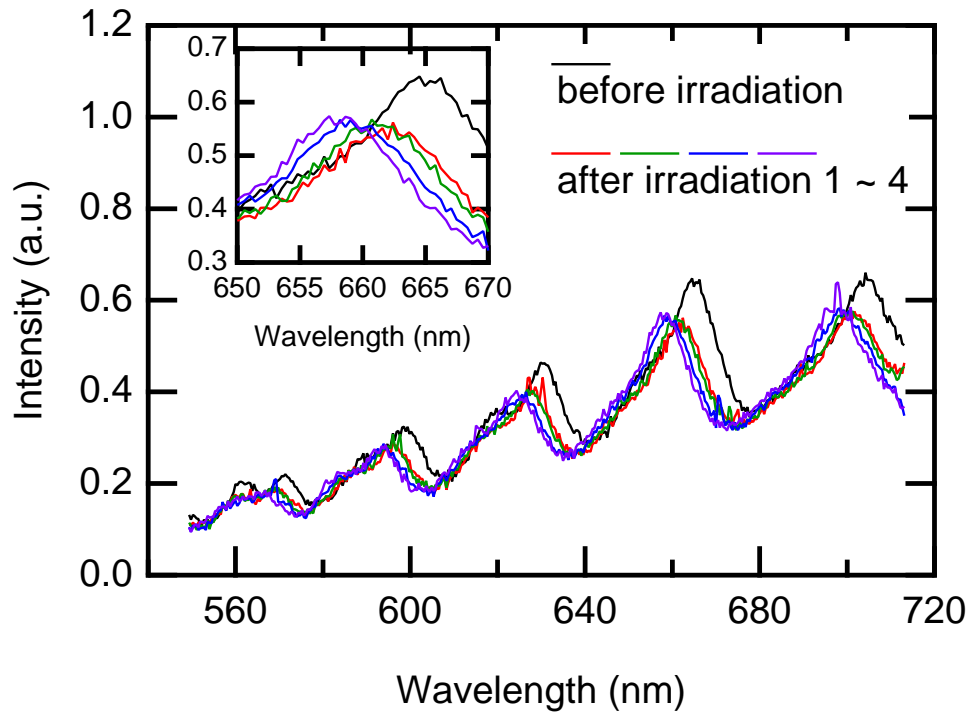}
\caption{Blue shift of the white light scattering spectrum from a single levitated microsphere after repeated CO2 laser irradiations. The inset shows an expansion of the peak around 660 nm.}
\label{fig:tuning}
\end{figure}

\begin{figure}[htbp]
\centering
\includegraphics[width=0.8\linewidth]{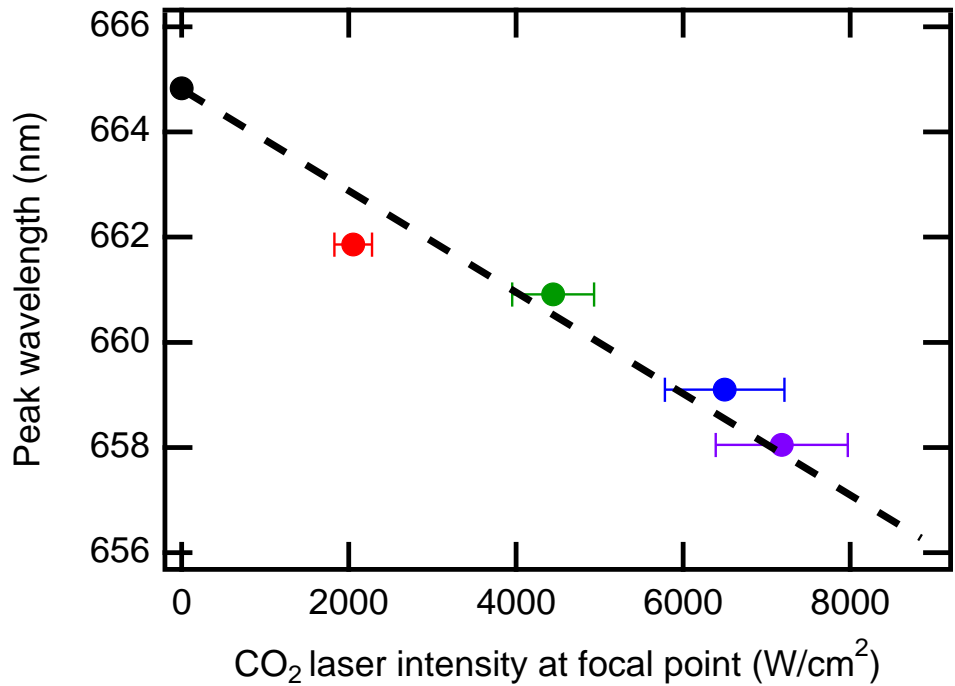}
\caption{Peak wavelength shift after each CO2 laser irradiation. The color of the data points correspond to the color in Fig. \ref{fig:tuning}}
\label{fig:peakshift}
\end{figure}

After the CO2 laser irradiation with the power of \SI{2}{kW/cm^2}, the white light scattering spectrum exhibited a slight blue shift of $\sim$\SI{3}{nm} as shown in Fig. \ref{fig:tuning} (red curve). The white light scattering spectrum continued to shift to shorter wavelength with increasing the CO2 laser power. Figure \ref{fig:peakshift}
summarizes the blue shift of the peak around \SI{660}{nm} as a function of the laser power. We achieved the WGMs tuning as large as \SI{7}{nm} via in-situ annealing. Note that very-little or no blued shift was observed, if we re-irradiated the levitated microsphere using the CO2 laser with the same power.

The temperature of the levitated microsphere during the annealing was determined by power balance among the CO2 laser heating, heat conduction through the surrounding gas, and thermal radiation. Here, we assume that the temperature distribution is homogeneous within the microsphere. Although this assumption seems unrealistic\cite{millen_nanoscale_2014}, the result gives us a rough estimation of the microsphere's steady state temperature.

Heat loss due to natural convection is negligible in our case, since the system's Grashof Number\cite{mahony_heat_1957} is very small owing to the small size of the microspheres (the radius $r=$ \SI{1.5}{\micro \meter}). Using kinematic viscosity and thermal expansion coefficient of air, \SI{1.5e-5}{m^2/s}\cite{dwight_e._gray_ed._american_1957} and \SI{3.6e-3}{K^{-1}}\cite{roebuck_joule-thomson_1930}, the Grashof number is calculated to be $G=5.3\Delta T\times 10^{-10}$, where  $\Delta T$ is the temperature difference between the heated microsphere and the air without heating. Then, the natural convective heat loss can be neglected compared with the conductive heat loss\cite{mahony_heat_1957} when the distance from the microsphere's surface is less than $r \times 1/\sqrt{G}$. Even if we take impractically large value $\Delta T = $\SI{10000}{K}, the critical distance is enough large $\sim$\SI{700}{\micro \meter}, which justifies the omission of the natural convection effect for the following calculation. 

Since the size of the microsphere is of the same order as the wavelength of the CO2 laser, Mie scattering theory\cite{bohren_absorption_1983} enables us to estimate the absorption cross section\cite{leinonen_python_????} from the extinction coefficient of the silica. As the value of the extinction coefficient of silica ranges from 0.02 to 0.5 depending on the sample\cite{kitamura_optical_2007}, we use the typical value $\kappa = 0.2$ for the calculation, which gives the absorption cross section $C_\mathrm{abs}\sim$\SI{3.2}{\micro \meter^2}. The laser heating power is
\[
\dot{q}_\mathrm{laser}=C_\mathrm{abs}I_\mathrm{CO2},
\]
where $I_\mathrm{CO2}$ is the CO2 laser intensity at the surface of the microspheres.

We assumed that the heat conduction through the surrounding gas occurs in the continuum regime, as the system's Knudsen number is $\sim 0.02$. The conductive cooling power is calculated to be\cite{liu_heat_2006}
\[
\dot{q}_\mathrm{air}=4\pi r\int_{T_g}^{T_p}k_g dT,
\] 
where $k_g$ is the temperature dependent heat conduction coefficient and  $T_p$ ($T_g$) is the temperature of the microspheres (the surrounding gas).

We calculate the radiative heat loss also from Mie scattering theory\cite{bohren_absorption_1983,liu_theoretical_2006} using frequency dependent complex refractive index and absorption cross section\cite{kitamura_optical_2007} as
\begin{align*}
\dot{q}_\mathrm{rad}&=f(T_p)-f(T_g),\\
f(T&)=f(T)=\int_0^\infty d\omega  4\pi \mathcal{P}_e(T) C_\mathrm{abs}(\omega)\\
&=\int_0^\infty d\omega  \dfrac{\hbar\omega^3}{4\pi^3c^2}\dfrac{4\pi \mathcal{P}_e(T) C_\mathrm{abs}(\omega)}{\exp\left(\hbar \omega/k_B T \right)-1},
\end{align*}
where $\mathcal{P}_e(T)$ is Plank distribution. The resulting cooling power is three orders of magnitude smaller than the conductive cooling power at the temperature less than \SI{1000}{K}. Therefore, we neglect the radiative cooling process here.

The steady state temperature of the levitated microsphere under CO2 laser irradiation is determined by numerically solving the equation $\dot{q}_\mathrm{laser}=\dot{q}_\mathrm{air}$. With the maximum CO2 laser intensity in our experiment, $I_\mathrm{CO2}=$\SI{7180}{W/cm^2}, we achieve the temperature $\sim$\SI{600}{K}. After the annealing at this temperature, we expect a few percentage reduction of the radius of the microsphere due to the removal of water molecules and organic impurities from the interior of micropores and the surface of the silica microspheres\cite{cho_stabilization_2014}. This shrinkage is consistent with the observed \SI{7}{nm} blue shift of the WGM resonances ($\sim$ 1 percent change of the wavelength).

\section{Conclusion}
This study demonstrated the in-situ tuning of WGM resonances of optically levitated microspheres through the annealing process. The thermal balance between CO2 laser heating and the heat conduction into the surrounding air determines the annealing temperature. With changing temperature, we can achieve the different WGM resonances because of the shrinkage of the microspheres.  

The rapidly growing field of the levitated optomechanics requires the diversity of the material for the levitated particles\cite{neukirch_multi-dimensional_2015,rahman_burning_2016,juan_near-field_2016}, as the internal degree of freedom in the material could couple with the optomechanical interaction and will open up the new protocol for the quantum state control of the levitated particles\cite{yin_large_2013}. Our results suggest the possibility of adjusting the material properties finely and even chemically modifying the particles during the levitation. In-situ fine control of the levitated particle's properties in vacuum need further research, which would result in the melting and vaporization of the levitated particles. Further selective doping with specific dopants is also possible, which will expand the available internal degree of freedom of the levitated particles.

\textbf{Funding Information.} JSPS KAKENHI Grant Number JP15K13501, JP16H06505, JP16H03884.; The Murata Science Foundation.; the Izumi Science and Technology Foundation.

% Bibliography
\bibliography{00WGMtuning}

\end{document}